\documentclass[singlecol]{epl2}
\usepackage{graphicx}
\usepackage[all]{xy}
\usepackage{amsmath}
\usepackage{amssymb}
\title{Scalar-Kinetic Branes}
\shorttitle{Scalar-Kinetic Branes} %Insert here a short version of the title if it exceeds 70 characters

\author{Yu-Xiao Liu\inst{1} \and Yuan Zhong
\inst{1} \and Ke Yang\inst{1}} \shortauthor{Y.-X. Liu \etal}

\institute{
  \inst{1} Institute of Theoretical Physics,
             Lanzhou University, Lanzhou 730000,
             People's Republic of China
} \pacs{11.27.+d}{Extended classical solutions; cosmic strings, domain walls, texture}
\pacs{11.25.-w}{Strings and branes}
\pacs{04.50.-h}{Higher-dimensional gravity and other theories of gravity}

\abstract{This work tries to find out thick brane solutions in braneworld
scenarios described by a real scalar field in the presence of a
scalar-kinetic term $F(X,\phi)=X\phi^m$ with a single extra
dimension, where $X=\frac12\nabla_M\phi\nabla^M\phi$ stands for the
standard kinetic term and $m=0,1,2\cdots$. We mainly consider bent
branes, namely de Sitter and Anti-de Sitter four-dimensional slices.
The solutions of a flat brane are obtained when taking the
four-dimensional cosmological constant $\Lambda_4\rightarrow 0$.
When the parameter $m=0$, these solutions turn to those of the
standard scenario. The localization and spectrum of graviton on
these branes are also analyzed.}

\begin{document}

\maketitle

%\section{Section title}
%See fig.~\ref{fig.1}, table~\ref{tab.1} and eq.~(\ref{eq.1}).
%See also~\cite{b.a,b.b}.
%\begin{equation}
%\label{eq.1}
%0\neq1
%\end{equation}

%\begin{figure}
%\onefigure{epl-template.eps}
%\caption{Figure caption.}
%\label{fig.1}
%\end{figure}

%\begin{table}
%\caption{Table caption.}
%\label{tab.1}
%\begin{center}
%\begin{tabular}{lcr} first  & table & row\\
%second & table & row
%\end{tabular}
%\end{center}
%\end{table}

\section{Introduction}
In the past decade, the braneworld scenario has attracted a lot of interests {for}
it {gives an effective way to solve} the hierarchy problem by introducing two
3-branes which are embedded in a five-dimensional anti-de Sitter (AdS$_5$) spacetime
\cite{Randall:1999ee}. As another attractive property the Newtonian law of gravity
{with} a correction {is also be given} in this braneworld scenario
\cite{Randall:1999vf}. The branes in \cite{Randall:1999ee,Randall:1999vf} are set up
artificially, and the thickness of each brane is neglected. However, in superstring
theory, there seems to exist a fundamental length scale, so the thin brane model
should be modified.

In more realistic models
\cite{GW,Freedman2000,gell,GremmPRD,C.Csaki2000,Kobayashi2002} {branes with
thickness were} realized and fixed by introducing some bulk fields. {In the simplest
case a single scalar field with standard dynamics was investigated, and the Lagrange
density of {the scalar field} is given by
\begin{eqnarray}
  \mathcal{L}=X-V(\phi),
  \label{standard-L}
\end{eqnarray}
where $X=\frac12\nabla_M\phi\nabla^M\phi$ stands for the kinetic
term, and $V(\phi)$ is the potential. Therefore in the standard
thick braneworld scenario, the five-dimensional action reads
\begin{eqnarray}
  S=\int d^4xdy\sqrt
  g\left(-\frac{1}{4}R+\frac12\nabla_M\phi\nabla^M\phi-V(\phi)
  \right).
  \label{standard-action}
\end{eqnarray}}

Recently, {there are many works} concerning the thick braneworld scenario under
various gravity theories. In addition to the traditional Einstein's gravity theory
\cite{Freedman2000,GremmPRD,Wang2002,0601069v3}, there are works on thick brane
scenario {considered in} Weyl geometry theory
\cite{ThickBraneWeyl,Liu0708,Liu0803,yangke}, or in $f(R)$ gravity theory
\cite{f(R)-brane} (for a review of the $f(R)$ gravity see \cite{f(R)gravity}).

Even under the traditional frame of gravity theory, braneworld {scenarios have been
developed} along different directions. Symmetric and asymmetric single or multiple
branes were investigated in
\cite{AsymmetricBranes,doubleBranes,LiuJCAP2009,0907.1640,Liu2010GRS}. Some
physicists prefer to consider this scenario in higher dimension and multi-scalar,
gauge field or vertex backgrounds as shown in
\cite{ThickBraneDzhunushaliev,Parameswaran0608074,LiuJHEP2007,Guerrero:2009ac,Almeida:2009jc}.
Some other recent works concerned {on field localization and resonances on deformed
branes \cite{Cruz0912,Cruz0912a}, and on spectra of field fluctuations in braneworld
models with broken bulk lorentz invariance\cite{Koroteev0901}}.

The majority of works we mentioned above mainly concerned flat branes,
namely four-dimensional Minkowski slices which embedded
in the bulk, and the dynamics of the bulk fields are always
considered to be standard.

For many aspects, we also want to know what would happen
if our spacetime is not flat but curved. The most familiar nonflat
spacetime geometries with maximum symmetry are de Sitter (dS) and
anti-de Sitter (AdS). Not just for they are also solutions
to Einstein equations, but for they allow nonzero cosmological
constants $\Lambda_4$ (for dS $\Lambda_4>0$ and for AdS $\Lambda_4<0$), thus they shed
a light for solving the problem of the evolution of our universe and
the puzzle of dark energy.

In a thick braneworld model with nonflat spacetime
geometry, Einstein equations are difficult to deal with.
However, according to the so called first order formalism
\cite{Freedman2000,first1,Townsend,BazeiaPLB2008,C.Csaki2000,gell},
one can reduce the second order Einstein equations to some
first order ones. With this formalism, some discussions
have been given on {bent thick branes}
\cite{GremmPRD,Wang2002,0601069v3}. For a comprehensive
review on thick brane solutions and related topics please
see Ref. \cite{ThickBraneReview}.

On the other hand, models with nonstandard dynamics of {background scalar} fields
are also {investigated} by a lot of physicists. $K$-field theory
\cite{K-fields,BazeiaPLB2008,0808.1815,SUSY-K} is a typical model with nonstandard
dynamic terms.
 In this theory the standard kinetic term is generalized to an arbitrary
function of $X$, as a consequence, the Lagrange density reads
\begin{eqnarray}
\mathcal{L}=F(X)-V(\phi).
\end{eqnarray}
$K$-field theory has already played a very important role in cosmology, {because} it
offers a mechanism for the early time inflation \cite{k-Inflation}. In another model
inspired and developed by string theorists~\cite{sen}, a nonstandard dynamics of the
scalar field was suggested in a way very similar to the Born-Infeld extension of
standard electrodynamics~\cite{Born-infeld}. In this case, the Lagrange density of
the background scalar field is given by
\begin{eqnarray}
\mathcal{L}=-V(\phi)\sqrt {1-2X}.\label{tachy-dy}
\end{eqnarray}
Models with Lagrange given by (\ref{tachy-dy}) or its
extensions have shown applications in D-brane and k-essence
theories~\cite{sen}. Obviously, when $X\ll 1$, the Lagrange
density~(\ref{tachy-dy}) reduces to
\begin{eqnarray}\label{smallX}
\mathcal{L}=-V(\phi)+V(\phi) X+{\mathcal {O}(X^2)}.
\end{eqnarray}
In the present paper, we extend the kinetic
term of the scalar field $X$ to a scalar-kinetic coupling form
$F(X,\phi)=W(\phi)X$, and then the Lagrange density takes the form
\begin{eqnarray}\label{S-K}
\mathcal{L}=-V(\phi)+W(\phi)X.
\end{eqnarray}
This model can be applied to describe systems with a static
background scalar field $\phi(y)$ which has a tiny
variation along only the fifth-dimension, i.e.,
$d\phi(y)/dy\approx0$, and so $\phi$ as well as $V(\phi)$
can be regarded as some constants. It is well known that in
the thin brane scenarios there is no bulk scalar field
involved, instead, a bulk cosmological constant is
introduced. On the other hand, if one takes the scalar
potential as a constant in the thick brane frame, then this
constant potential can be regarded as a bulk cosmological
constant. Thus when considered a small perturbation of the
constant scalar potential along the fifth-dimension, such
systems can be described by (\ref{S-K}).
 In principle, $W(\phi)$ can be any arbitrary
function of $\phi$, but for simplicity, we focus only on the case
$W(\phi)=\phi^m$ with $m=0,1,2\cdots$. We will give the
analytic solutions of our model in flat and dS/AdS 3-brane
cases. It turns out that the results of our model {are} consistent with
those of the standard kinetic model studied in~\cite{GremmPRD} when
take $m=0$.

The investigation is organized as follows: in the next section we will give a brief
description {of} our framework, and give the dynamic equations for general systems
governed by the Lagrange denstity $\mathcal{L}=F(X,\phi)-V(\phi)$. Then we come back
to the case of scalar-kinetic extension~(\ref{S-K}), and take $W(\phi)$ as
{$\phi^m$}, in this case the dynamical equations can be solved in a closed form. We
assume the scalar and the warp factor depend only on the extra dimension throughout
the following discussions. A particular definition of the mass in the de Sitter and
anti-de Sitter {spacetimes allows} us to discuss the stability of our model as well
as the effective gravity on the {branes}. We also obtain a correction of the
Newtonian potential between two massive points lying on the brane. It turns out that
the correction is different from the one given by RS model~\cite{Randall:1999vf}.

\section {The framework and the dynamic equations}
\label{secFramework}

A thick braneworld scenario with non-standard dynamics can
be described by a real scalar field $\phi$ coupled with
gravity via the Einstein-Hilbert action, which has the
general form
\begin{eqnarray}
  S=\int d^4xdy\sqrt g\left(-\frac{1}{4}R+\mathcal {L}(X,\phi)
  \right),
  \label{action}
\end{eqnarray}
 The space-time structures we are
interested in here are mainly dS$_4\subset$ AdS$_5$ and
AdS$_4\subset$ AdS$_5$ with the line elements \cite{Mannheim2005}
\begin{eqnarray}
  ds_5^2=e^{2A}\big(dt^2-e^{2H
  t}(dx^2_1+dx^2_2+dx^2_3)\big)-dy^2
\label{dS5}
\end{eqnarray}
and{\footnote{The AdS$_4$ parameter $H$ is
  related to the dS$_4$ one by $H$(AdS)$=-iH$(dS) since $x_3$(AdS)
   is equavalent to $it$(dS)}}
\begin{eqnarray}
  ds_5^2=e^{2A}\big(e^{2H
  x_3}(dt^2-dx^2_1-dx^2_2)-dx^2_3\big)-dy^2,
\label{AdS5}
\end{eqnarray}
respectively. Here $e^{2A}$ is the warp factor and {$H$ is the de Sitter or anti-de
Sitter parameter, which {is related to} the four-dimensional cosmological constant
by $\Lambda_4=\pm 3H^2$, the sign of the proportionality coefficient is determined
by the geometry of the spacetime. For de Sitter space $\Lambda_4=3H^2>0$, and for
anti-de Sitter space $\Lambda_4=-3H^2<0$.} The case of the Minkowski spacetime
($\Lambda_4=0$) is obtained in the limit $H\rightarrow 0$.

The scalar field dynamics is governed by the Lagrange density
\begin{eqnarray}
   \mathcal {L}= F(X,\phi)-V(\phi),
\end{eqnarray}
which is a little different from the one in \cite{0808.1815}, where $\mathcal {L}=
F(X)-V(\phi)$. This means that the coupling between the scalar field and its kinetic
term $X$ is allowed in our model. Explicit examples given behind show that {a series
of} analytical solutions can be obtained with this assumption.

The equation of motion for the scalar field takes the form
\cite{0808.1815}:
\begin{eqnarray}
  -4A'\phi'^2\mathcal {L}_X=(\mathcal {L}-2\mathcal {L}_XX)',
  \label{EOM1}
\end{eqnarray}
where the prime denotes the derivative with respect to the extra
dimension $y$, and we take the notation of $\mathcal
{L}_X=\frac{\partial\mathcal{L}}{\partial X}$. The scalar field is
supposed to be a function of $y$ only, i.e., $\phi=\phi(y)$.

In the case of the de Sitter metric (\ref{dS5}), we have nonzero
components of Christoffel symbols: $\Gamma^\alpha_{\beta
4}=A'\delta^\alpha_\beta$, $\Gamma^4_{0 0}=A'e^{2A}$, $\Gamma^4_{i
j}=-\delta_{i j}e^{2(H t+A)}A'$ and $\Gamma^i_{j 0}=\delta^i_j H$,
where $i, j=1, 2, 3$, while Greek letters $\alpha, \beta=0, 1, 2,
3$. The non-vanishing components of the Ricci tensor are
$R_{00}=-3H^2+4e^{2A}A'^2+e^{2A}A''$, $R_{ij}=-\delta_{ij}e^{2H
t}R_{00}$ and $R_{44}=-4(A'^2+A'')$. The components of the
energy-momentum tensor are given by $T_{00}=-e^{2A}\mathcal {L}$,
$T_{ij}=\delta_{ij}e^{2A+2H t}\mathcal{L}$ and
$T_{44}=\mathcal{L}-2X\mathcal {L}_X$. By variating the modified
action (\ref{action}) with respect to the metric (\ref{dS5}) we can
finally reach to the Einstein equations:

\begin{subequations}
\begin{eqnarray}
  A''+H^2 e^{-2A}&=&-\frac23\phi'^2\mathcal {L}_X, \label{dSEE1}\\
  A'^2-H^2 e^{-2A}&=&\frac13(\mathcal {L}-2X\mathcal{L}_X). \label{dSEE2}
  \end{eqnarray}
  \label{dSEE}
\end{subequations}

The equations corresponding to metric (\ref{AdS5}) {can be} obtained {by replacing
$H^2$ with $-H^2$}, thus in the discussions below we always consider the de Sitter
brane. Note that when $H=0$, these equations reduce to {the ones}
in~\cite{0808.1815}. On the other hand, the Einstein equations for models with a
nonvanishing parameter $H$ as well as a standard kinetic term $\mathcal{L}(X,\
\phi)=X-V(\phi)$ have been given in~\cite{GremmPRD,0601069v3}:
\begin{subequations}
\begin{eqnarray}
 A''+H^2 e^{-2A}&=&-\frac23\phi'^2,\\
 A'^2-H^2 e^{-2A}&=&\frac16\phi'^2-\frac13V(\phi),
\end{eqnarray}
\end{subequations}
which are consistent with (\ref{dSEE}).
Note that the dynamic
equations are not independent, one can easily prove that
(\ref{EOM1}) can be obtained from (\ref{dSEE}). Thus, solving
Einstein equations (\ref{dSEE}) is enough.

\section{The braneworld scenario}
\label{secBrane}

When {referring} to the problem of finding a thick braneworld solution, it means
{that} one must find out three quantities: $\phi(y)$, $A(y)$ and $V(\phi)$, which
solve the dynamic equations. However, we have only two independent equations. As a
result, one of the quantities we are searching for must be given at first. For
simplicity, we assume that the warp factor takes the form
\begin{eqnarray}
  A=\ln\left( \cos(by)\right).
  \label{A(y)}
\end{eqnarray}
The warp factor $e^{2A}$ corresponding to eq.~(\ref{A(y)}) has naked singularities
at $y=\tilde y=\pm\pi/(2b)$, where the scalar $\phi$ as well as the potential
$V(\phi)$ diverge. This feature seems contradicting to our assumption that
$V(\phi)\approx$const., however in the area around the brane, $V(\phi)$ varies
slowly so that our model is reasonable at least in the inner space. In fact, if we
regard the scalar field as a modulus from some {compactification} manifold, then
these singularities can indicate that the compactification manifold shrinks to zero
size or extends infinitely large, as a result, the five dimensional truncation
cannot be used here. In addition, such singularities are very similar to those
confronted in the AdS flow to $N=1$ super Yang-Mills theory~\cite{Girardello}, that
means it may be solved either by lifting the dimension to ten or by string theory.
There are works where singularities in five dimensions actually correspond to
non-singular ten dimensional geometries~\cite{9906194}.

For a dS$_4$ brane we usually expect a horizon at a finite
distance $y=\tilde y$, so the naked singularity can be
regarded as the horizon for the dS$_4$ brane. The argument
is not the same for the AdS$_4$ brane. In the case of a
flat brane, we have a brane interpolated between AdS$_5$
spaces which have regular horizons infinitely far away from
$y=0$. It is obvious that as $b\rightarrow 0$ we have
$\tilde y\rightarrow\infty$. Thus $b$ should be related to
the parameter $H$.

\subsection{Model and Solutions}
With the warp factor given by~(\ref{A(y)}), the solution of Einstein
equations (\ref{dSEE}) can be worked out in a closed form. We
consider the following extension of the standard kinetic term $X$:
\begin{eqnarray}
  F(X,\phi) = X \phi^m, \label{FXphi}
\end{eqnarray}
where $m=0,1,2,\cdots$. Einstein equation (\ref{dSEE1}) turns to
\begin{eqnarray}
\left(b^2-H^2 \right) \sec^2(b y)=\frac{2}{3} \phi ^m\phi'^2.
\label{phim}
\end{eqnarray} The solutions of this equation are slightly varied for different
$m$.

When $m=4n,\ n=0,1,2,\cdots$, {we get $4n+2$ solutions for
eq.~(\ref{phim})}, we study only one of them:
\begin{eqnarray}
\phi(y)=\big[(2n+1)\beta ~\textrm{arcsinh}(\tan (b
y))\big]^{\frac{1}{2n+1}}
   \label{phi1},
\end{eqnarray}
where we have set the integral constant to zero and $\beta$ is
defined as
\begin{eqnarray}
\beta=\frac{1}{b}\sqrt{\frac{3}{2}\left(b^2-H^2 \right)}.
\end{eqnarray}
From another Einstein equation (\ref{dSEE2}), $V(\phi)$ can be
solved as
\begin{eqnarray}
 V(\phi )=3 b^2-\frac{9}{4} \left(b^2-H^2 \right)
 \cosh ^2\left[\frac{\phi ^{2n+1}}{(2n+1)\beta}\right].
  \label{V(phi)1}
\end{eqnarray}
{Note that $\phi(y)$ and $V(\phi)$ are divergent at the boundary $y=\pm\pi/(2b)$ and
$\phi=\phi(\pm\pi/(2b))=\pm\infty$, respectively.}

When $m=4n+2,\ n=0,1,2,\cdots,$ we will find that
\begin{eqnarray}
  \phi(y)&=& \big[(2n+2)\beta
          ~\textrm{arcsinh}(\tan (b y))\big]^{\frac{1}{2n+2}}
          \label{phi2},\\
 V(\phi )&=& 3 b^2-\frac{9}{4} \left(b^2-H^2 \right)
 \cosh ^2\left[\frac{\phi ^{2n+2}}{(2n+2)\beta}\right] , \label{Vphi2}
\end{eqnarray}
where $\phi(y)$ is invalid for negative $y$.

For odd $m=2n+1,\ n=0,1,2,\cdots$, the solution is
\begin{eqnarray}
  \phi(y)=\left[\frac{(2n+1)}{2}\beta
          ~\textrm{arcsinh}(\tan (b y))\right]^{\frac{2}{(2n+1)}},\\
 V(\phi )=3 b^2-\frac{9}{4} \left(b^2-H^2 \right) \cosh ^2
          \left[\frac{2\phi ^{\frac{2n+1}2}}{(2n+1 )\beta}\right]. \label{Vphi3}
\end{eqnarray}
Obviously $V(\phi)$ in (\ref{Vphi3}) is invalid for negative $\phi$.
One can see that if $X$ is not coupled with $\phi^{4n}$, but with
$\phi^{4n+2}$ or $\phi^{2n+1}$, some problems appear. Therefore,
taking $F(X,\phi) = X \phi^{4n}$ is the best choice.

{In the standard dynamic case ($m=0$), i.e., $L(X,\phi)=X-V$, the solution is then
given by}
\begin{eqnarray}
  A&=&\ln[\cos(b y)], \nonumber\\
  \phi _0(y)&=&\beta~\textrm{arcsinh}(\tan (b y)),\\
 V_0(\phi )&=&3 b^2-\frac{9}{4} (b^2 -H^2)
 \cosh ^2({\phi }/{\beta}).\nonumber
\end{eqnarray}
One can find that the above solution is the same one given in~\cite{GremmPRD} if one
{notices} that expressions below are actually identical:
\begin{eqnarray}
  \textrm{arcsinh}(\tan (b y)) %&=&\text{EllipticF}[by,\ 1]\nonumber\\
  %&=&2 arcsinh(\tan(by/2)) \\
  &=&\ln\left(\frac{1+\tan(by/2)}{1-\tan(by/2)}\right).
\end{eqnarray}

For a four-dimensional dS slice, the only constrain here is
$0<H^2<b^2$. On the other hand when reversing the sign of
$H^2$, we get solutions to the AdS$_4\subset$ AdS$_5$
space-time without constrain.

%%%%%%%%%%%%%%%%%%%%%%%%%%%%%%%%%%%%%
\subsection{Stability of the {model}}

In our {model}, modifications of the scalar field dynamics are introduced, thus it
is important to {know} if these modifications will contribute to destabilize the
geometric degrees of freedom of the braneworld model.

As illustrated in~\cite{gell} the fluctuations of both the metric
and the scalar field need to be considered. However, the general
treatment is difficult, since the fluctuations of the scalar field
will couple with the metric fluctuations, yet, one can continue the
discussion in a sector given in~\cite{Freedman2000}, where the
metric fluctuations decouple from the scalar and satisfy a simple
wave equation.

{Following~\cite{Wang2002}, we consider the gravity
perturbations:
\begin{equation}\label{comformal}
 ds^2=e^{2 A(z)}\left[(g_{\mu\nu}+h_{\mu\nu})dx^\mu dx^\nu-dz^2\right],
\end{equation}
 where
$g_{\mu\nu}=g_{\mu\nu}(x)$ represents the four-dimensional AdS,
Minkowski, or dS metric, $h_{\mu\nu}=h_{\mu\nu}(x,z)$ is the metric
perturbation which satisfies the transverse-traceless (TT)
condition, i.e.,
\begin{equation}
h^\lambda_{~~\lambda}=0=h_{\mu\nu;\lambda}g^{\nu\lambda}.\label{TT}
\end{equation}
{Notice} we have used a coordinate transformation
\begin{eqnarray}
  z(y)&=&\int e^{-A(y)}dy =\int   \sec (by) dy\nonumber\\
  %&=&\frac{2 arcsinh\left(\tan \left(\frac{b y}{2}\right)\right)}{b}
  &=&\frac{1}{b} arcsinh\left(\tan (b y)\right)
  \label{relation z y}
\end{eqnarray}
to rewrite the metrics~(\ref{dS5}) and~(\ref{AdS5}) into a conformal form.
Then it follows the main equation for $h_{\mu\nu}$~\cite{Kobayashi2002}:
\begin{equation}
 h''_{\mu\nu}+3A'h'_{\mu\nu}-\Box h_{\mu\nu}-2\alpha^2h_{\mu\nu}=0.
\end{equation}
{Here the prime denotes the derivative with respect to the extra dimension $z$.} We
define the mass by the last two terms of the equation above as
\begin{equation}\label{mass}
 -m^2h_{\mu\nu}=\Box h_{\mu\nu}-2\alpha^2h_{\mu\nu}.
\end{equation}
Then by introducing a polarization tensor
$\epsilon_{\mu\nu}(x^\alpha)$ which depends only on the
four-dimensional coordinates $x^\alpha$ and satisfies the TT
condition~(\ref{TT}), and taking the decomposition of $h_{\mu\nu}$
\begin{equation}
h_{\mu\nu}=e^{-3A(z)/2}\epsilon_{\mu\nu}(x^\alpha)\psi(z),
\end{equation}
one obtains a schrodinger like equation for $\psi(z)$
\begin{equation}
\label{sch}
 \left[-\frac{d^2}{dz^2}+V_{Sch}(z)\right]\psi(z)=m^2{\psi(z)}
\end{equation}
with
\begin{equation}
 V_{Sch}(z)=\frac94 A^{\prime 2}(z)+\frac32 A^{\prime\prime}(z).
\label{V(z)}
\end{equation}
In $z$-coordinate, $A(z)=-\ln[\cosh (b z)]$, thus
\begin{eqnarray}
  V_{Sch}=\frac{9}{4} b^2 -\frac{15}{4} b^2 \text{sech}^2(b
  z).
  \label{V}
\end{eqnarray}
This is the well known P\"oschl-Teller potential which appears in many
works~\cite{GremmPRD,Wang2002,0601069v3,LiuJCAP2009,PTPotential,Liu0803}. The
potential here supports two bound states: $\psi_0(z)\sim\cosh^{-3/2}(bz)$ and
$\psi_1(z)\sim{\sinh(bz)}{\cosh^{-3/2}(bz)}$ with eigenvalues $m^2_0=0$ and
$m_1^2=2b^2<(9/4) b^2$, respectively. {Furthermore}, there is also a continuum of
states which asymptote to plane waves when far away from the brane.

The spectrum of the KK modes is essential for us to discuss the effective gravity on
the brane. We have shown that with the mass defined by~(\ref{mass}), there are a
normalizable zero mode which will give rise to 4D gravity on the 3-brane, and a
{separate and bound} massive mode as well as a continuum {of} massive KK modes. We
will show that it is {the continuous KK modes} that give a correction to the
Newtonian potential, while the {bound} massive mode has no contribution to the
correction.

However, in some other literatures such as~\cite{GremmPRD}, the mass is defined
differently, as a consequence, there is no zero mode for the AdS brane case, strange
still, for the dS brane there is a tachyonic state which may show a sign of
instability. If we follow the definition of mass in~\cite{GremmPRD}, then the
correction of the 4D gravity in the case of de Sitter and AdS case can no longer be
treated under the traditional methods. For all {of} the reasons, we take the mass
defined by~(\ref{mass}).

Now let us turn to the discussion of the effective gravity
on the brane. For simplicity, we consider two point-like
sources with mass $M_1$ and $M_2$, both confined at the
origin of the fifth-dimension, i.e., $z=0$. This assumption
is justified when the thickness of the brane is small as
compared with the bulk curvature. Then the effective
potential is given by~\cite{C.Csaki2000}:
\begin{equation}
U(r)=
G_N\frac{M_1M_2}{r}+\frac{M_1M_2}{M_*^3}\int_{m_0}^{\infty}{dm
\frac{e^{-mr}}{r}|\psi_m(0)|^2},
\label{Newtonian_potential}
\end{equation}
where $G_N=M_4^{-2}$ is the four-dimensional coupling constant, i.e., the Newton's
gravitational constant, $M_*$ is the fundamental five-dimensional Planck scale, and
$m_0=3b/2$ is the minimal eigenvalue where the continuous KK modes start at. It is
the massless zero mode which offers the four-dimensional Newtonial interaction
potential, {and all the} massive KK modes give a correction to the standard
Newtonian potential, however since the wave function of the first excited state
$\psi_1(z)$ vanishes at $z=0$, we start the integral from $m_0=3b/2$. With a
coordinate transformation $l=bz$, the Shr\"odinger like equation~(\ref{sch}) turns
into
\begin{equation}
-\psi''(l)-\frac{15}{4}\text{sech}^2(l)\psi(l)=M^2\psi(l),
\label{KK_Schrodinger_Eqs_l}
\end{equation}
where the prime denotes the derivative with respect to the
coordinate $l$, and $M=\sqrt{\frac{m^2}{b^2}-\frac{9}{4}}$.
The solution of this equation is given by a linear
combination of the Legendre functions:
\begin{equation}
\psi_m(l)=C_1~\text{P}\left(\frac32, i M,
\tanh(l)\right)+C_2~\text{Q}\left(\frac32, i M,
\tanh(l)\right), \label{KK_modes_l}
\end{equation}
where $C_1$, $C_2$ are $M$-dependent parameters:
\begin{eqnarray}
C_1&=&\frac{\Gamma (1-i M)}{2}\nonumber \\
&+&\frac{\cosh(M \pi )^2\Gamma \left(-\frac{3}{2}-i M\right) \Gamma \left(\frac{5}{2}-i M\right)\Gamma (1+i M)}{2 \pi },\nonumber\\
C_2&=&\frac{M \Gamma \left(-\frac{3}{2}-i M\right) \Gamma
\left(\frac{5}{2}-i M\right) \Gamma (i M) \sinh[2 M \pi ]}{2 \pi
^2}.\nonumber
\end{eqnarray}
{Thus, by substituting $\psi_m(0)$ into eq.~(\ref{Newtonian_potential}),} we obtain
the correction of the Newtonian potential. It was found \cite{yangke} that the
correction of the Newtonian potential given by a P\"oschl-Teller potential is
\begin{equation}
\Delta U(r)\propto\frac{e^{-3b r/2 }}{M_*^3}\frac{M_1M_2}{r^2},
\end{equation}
which is different to the correction given by a
volcano-like potential~\cite{Randall:1999vf} where $\Delta
U(r)\propto1/r^3$, for more details please
see~\cite{yangke}.

\section{{Conclusions}}\label{secConclusion}

{In this work, we have studied a brane model in which the
standard kinetic term $X$ is extended to scalar-kinetic coupling
terms. Such extension can be regarded as a result of trying to
combine features of the K-field theory and the Born-Infeld extension
of the kinetic term under the condition $X\ll 1$. It is well known
that when $X\ll 1$ we have $\phi\approx$const. and
$V(\phi)\approx$const., while the constant scalar potential can be
identified with the bulk cosmological constant. If we introduce a
tiny perturbation to $V(\phi)$, then the scalar-kinetic coupling can
be applied to describe such system.

Another assumption of our model is the form of the warp factor which is given
by~(\ref{A(y)}). The warp factor implies a naked singularity at $y=\tilde
y=\pm\pi/(2b)$ where {the scalar field is} divergent. However, such singularity can
be solved by lifting the dimensions of spacetime to ten, or by using string theory.
The parameter $b$ relates to the de Sitter parameter $H$ which controls the
thickness of the brane: a finite and positive $b$ leads to finite $\tilde y$ which
implies a finite boundary or horizon of the de Sitter spacetime. As $b\rightarrow 0$
the horizon appears {at $y\rightarrow \pm\infty$}.

On the stability of the metric fluctuations we took the
definition of mass as in~\cite{Wang2002,Kobayashi2002}.
Then fluctuations of metric lead to a shr\"odinger equation
with the so called P\"oschl-Teller potential. This
potential supports two bound states along with a continuum
of states which asymptote to plane waves at infinity. With
this mass spectrum, we analyzed the effective gravity on
the brane. It is the zero mode which gives rise to the
standard Newtonian potential, and a correction is given by
all the massive modes. Although we have another bound state
$\psi_1(z)$ with $m=\sqrt 2b$, we showed that $\psi_1(z)$
has no contribution to the effective gravity, since
$\psi_1(0)=0$. Thus the continuous massive modes (separated
from the zero mode by a mass gap $\Delta m=3b/2$) offer the
Newtonian potential a correction: $\Delta
U(r)\propto\frac{e^{-3b r/2 }}{M_*^3}\frac{M_1M_2}{r^2}$,
which is much different to the one given by a volcano-like
potential.}

 This scalar-kinetic field may
have some applications in inflation, we hope to have some related
works reported in the future.

\acknowledgments This work was supported by the Program for New Century Excellent
Talents in University, the National Natural Science Foundation of China (No.
10705013), the Doctoral Program Foundation of Institutions of Higher Education of
China (No. 20090211110028), the Key Project of Chinese Ministry of Education (No.
109153), the Ying-Dong Huo Education Foundation of Chinese Ministry of Education
(No. 121106), and the Fundamental Research Funds for the Central Universities (No.
lzujbky-2009-54).

\end{document}